\documentclass[aps,prl,noshowpacs,twocolumn,reprint,amsmath,amssymb,superscriptaddress,longtable]{revtex4-1}

\usepackage{graphicx}
\usepackage{dcolumn}
\usepackage{bm}
\usepackage{tablefootnote}

\usepackage{amsthm}
\usepackage{mathtools}
\usepackage{amsmath}
\usepackage{amssymb}
\usepackage{graphicx}
\usepackage{url}
\usepackage{float}
\usepackage{bm}
\usepackage{ifthen}
\usepackage[usenames,dvipsnames]{color}
\usepackage{mathrsfs}
\usepackage[colorlinks=true,citecolor=blue,urlcolor=black]{hyperref}
\usepackage{float}
\usepackage{subfigure}
\usepackage{enumitem}
\usepackage{color}
\usepackage{setspace}
\usepackage{braket}
\usepackage{comment}
\usepackage{lipsum}
\usepackage{color}

\newcommand{\be}{\begin{equation}}
	\newcommand{\ee}{\end{equation}}

\begin{document}
	
	\title{Proof-of-Principle Demonstration of Fully-Passive Quantum Key Distribution}

	\author{Chengqiu Hu}
	\thanks{hucq@hku.hk}
	\affiliation{Department of Physics, University of Hong Kong, Pokfulam Road, Hong Kong}
	
	\author{Wenyuan Wang}
	\affiliation{Department of Physics, University of Hong Kong, Pokfulam Road, Hong Kong}
	
	\author{Kai-Sum Chan}
	\affiliation{Department of Physics, University of Hong Kong, Pokfulam Road, Hong Kong}
	\affiliation{Quantum Bridge Technologies, Inc., 100 College Street, Toronto, ON M5G 1L5, Canada}
	
	\author{Zhenghan Yuan}
	\affiliation{Department of Physics, University of Hong Kong, Pokfulam Road, Hong Kong}

	\author{Hoi-Kwong Lo}
	\thanks{hoikwong@hku.hk}
	\affiliation{Department of Physics, University of Hong Kong, Pokfulam Road, Hong Kong}
	\affiliation{Quantum Bridge Technologies, Inc., 100 College Street, Toronto, ON M5G 1L5, Canada}
	\affiliation{Centre for Quantum Information and Quantum Control (CQIQC), Dept. of Electrical \& Computer Engineering, University of Toronto, Toronto,  Ontario, M5S 3G4, Canada}
	\affiliation{Centre for Quantum Information and Quantum Control (CQIQC), Dept. of Physics, University of Toronto, Toronto,  Ontario, M5S 3G4, Canada}

	\begin{abstract}
		Quantum key distribution (QKD) offers information-theoretic security based on the fundamental laws of physics. However, device imperfections, such as those in active modulators, may introduce side-channel leakage, thus compromising practical security. Attempts to remove active modulation, including passive decoy intensities preparation and polarization encoding, have faced theoretical constraints and inadequate security verification, thus hindering the achievement of a fully passive QKD scheme. Recent research\cite{wang2022fully, zapatero2023fully} has systematically analyzed the security of a fully passive modulation protocol. Based on this, we utilize the gain-switching technique in combination with the post-selection scheme and perform a proof-of-principle demonstration of a fully passive quantum key distribution with polarization encoding at channel losses of 7.2 dB, 11.6 dB, and 16.7 dB. Our work demonstrates the feasibility of active-modulation-free QKD in polarization-encoded systems.
		
	\end{abstract}

	
	\date{\today}
	\maketitle

	\textbf{Introduction.} Quantum Key Distribution (QKD) has emerged as a promising technology to ensure information-theoretically secure communications\cite{ch1984quantum, ekert1991quantum}. Despite its strong theoretical foundation, practical QKD systems are susceptible to side-channel attacks, which can compromise their practical security. While measurement-device-independent (MDI) QKD\cite{PhysRevLett.108.130503, liu2019experimental,cao2020long} and twin-field (TF) QKD protocols\cite{lucamarini2018overcoming, wang2022twin, liu2023experimental, zhou2023twin} have been proposed to address side-channel vulnerabilities in the measurement unit, source modulators in current implementations still present significant challenges\cite{lutkenhaus2002quantum, pang2020hacking}. 

The conventional method of active modulation using phase and intensity modulators can introduce side channels and are vulnerable to Trojan Horse attacks,\cite{gisin2006trojan, tamaki2016decoy, bourassa2022measurement, yoshino2018quantum, huang2019laser, PhysRevA.88.022308} in which an eavesdropper may inject strong light into the modulator and analyze the back-reflected signal to obtain information. In addition, active modulators face other practical issues. The pattern effect arises in high-speed systems when the modulation strength of adjacent pulses influences each other, causing undesired correlations\cite{roberts2018patterning, lu2021intensity}. The commonly used electro-optic modulators often require high driving voltages, posing technical challenges for high-speed electronic systems. Besides, commercially available electro-optic modulators tend to be expensive, bulky, and difficult to integrate\cite{paraiso2019modulator, woodward2021gigahertz}.

Previous research has attempted to implement passively decoy state generating or polarization encoding\cite{curty2009non, curty2010passive, sun2014experimental, sun2016experimental, zhang2012experimental, li2022passive}. These approaches have paved the way for further innovation, though they do not achieve a complete passive modulation. Another approach involves the use of injection locking and direct phase modulation\cite{comandar2016quantum, PhysRevX.6.031044}, which enables modulator-free QKD systems based on phase encoding, such as time-bin encoding. However, direct phase modulation often requires meticulous fine-tuning and may involve considerable technical complexity. In recent work\cite{wang2022fully}, a fully passive QKD solution was proposed along with a systematic security analysis, but so far a comprehensive experimental demonstration is yet to be completed.

	\begin{figure}[b]
		\includegraphics[scale=0.80]{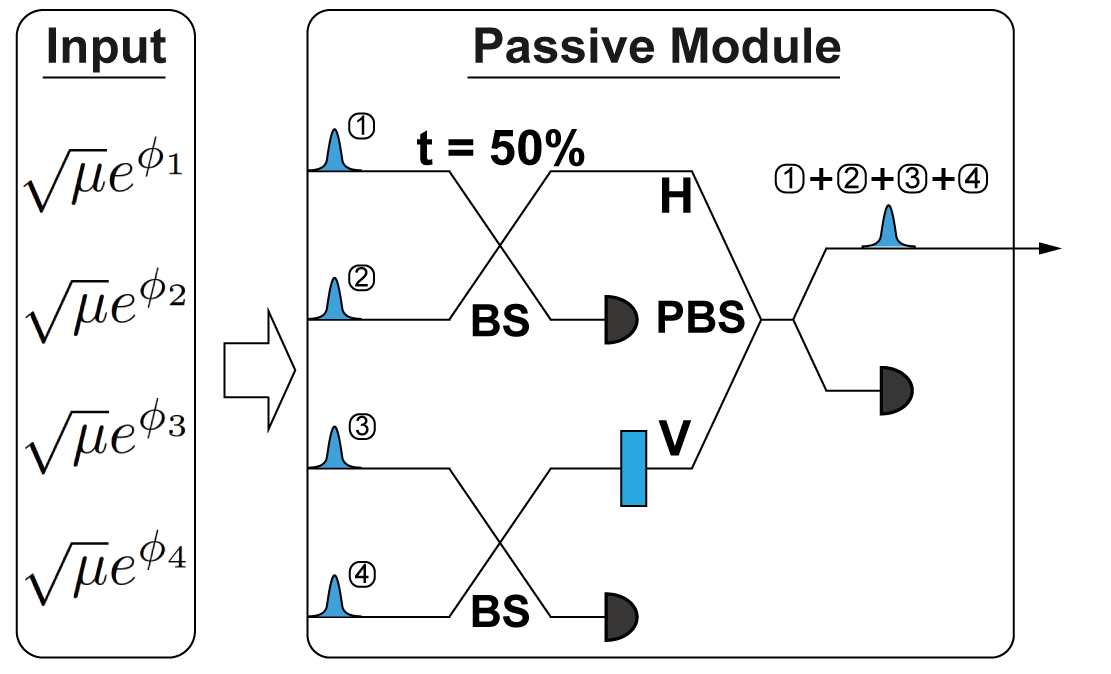}
		\caption{\textbf{Illustration of the fully passive protocol.} The input here consists of four coherent light pulses with random phases. The so-called passive module is used for fully passive decoy-state preparation and polarization encoding.}
		\label{fig:1}
	\end{figure}
	\begin{figure*}[htb]
		\includegraphics[scale=0.98]{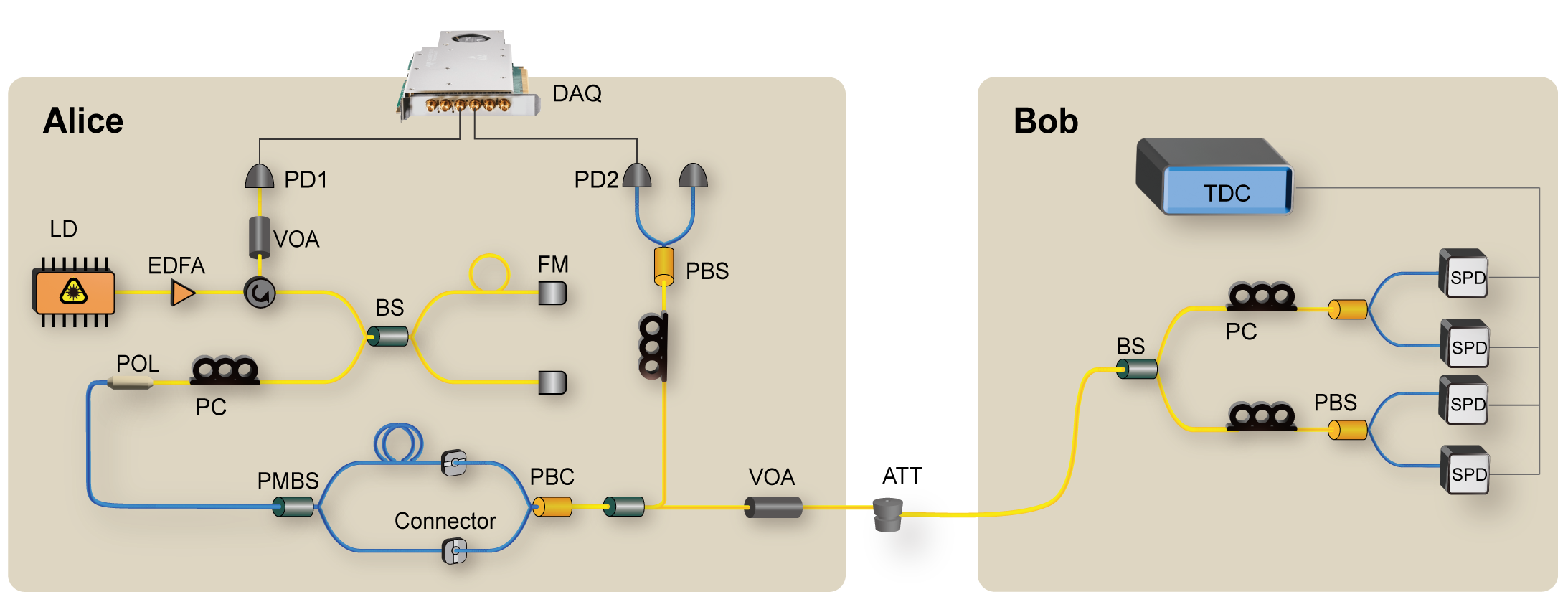}
		\caption{\textbf{Experimental setup.} On the Alice side, we use an unequal arm MZI for passive decoy-state preparation and employ a polarization beam splitter and a polarization combiner for polarization synthesis. Detector PD1 is used for local measurements on the Z basis, while detector PD2 is used for local measurements on the X basis. The results are collected by a data acquisition card (DAQ).  LD: laser, EDFA: erbium-doped fiber amplifier, VOA: variable optical attenuator, PD: photodiode, DAQ: data acquisition card, BS: beamsplitter, FM: Faraday mirror, PC: polarization controller, POL: polarizer, PMBS: polarization-maintaining beamsplitter, PBC (PBS): polarization beam combiner (splitter), ATT: attenuator. On the Bob side, a standard BB84 decoding setup with four single-photon detectors is used. The data is collected by a time-digital converter. SPD: single-photon detector, TDC: time-digital convertor. Single-mode fiber is represented by yellow in this figure while polarization-maintaining fiber is represented by blue. }
		\label{fig:2}
	\end{figure*}
The implementation of fully passive QKD presents several major challenges. First, the scheme requires input coherent pulses with independent and random phases, while ensuring that these pulses have highly consistent intensity and frequency to achieve sufficient interference visibility. Our solution involves using a single laser instead of multiple lasers, driving it in a gain-switch manner\cite{yuan2014robust} to generate intrinsic randomness. By introducing a delay between consecutive pulses, they can be treated as independent pulse sources, thereby avoiding the frequency deviation associated with multi-laser protocol. Another challenge here is that the fully passive QKD schemes require local measurements to collect large amounts of data due to post-selection needs. We address this issue by using a clever experimental setup, requiring a minimum of only two photodetectors to complete local measurements, and utilizing a high-speed acquisition card for continuous data collection. By adeptly utilizing a customized post-selection approach, we efficiently tackle the challenges posed by the signal intensity and polarization coupling in a fully passive setup. Our proof-of-principle experiment validates that this protocol\cite{wang2022fully} can perform passive decoy-state preparation and polarization encoding simultaneously, successfully demonstrating a fully passive QKD system with improved security and robustness.

	\begin{figure*}[htb]
		\includegraphics[scale=0.9]{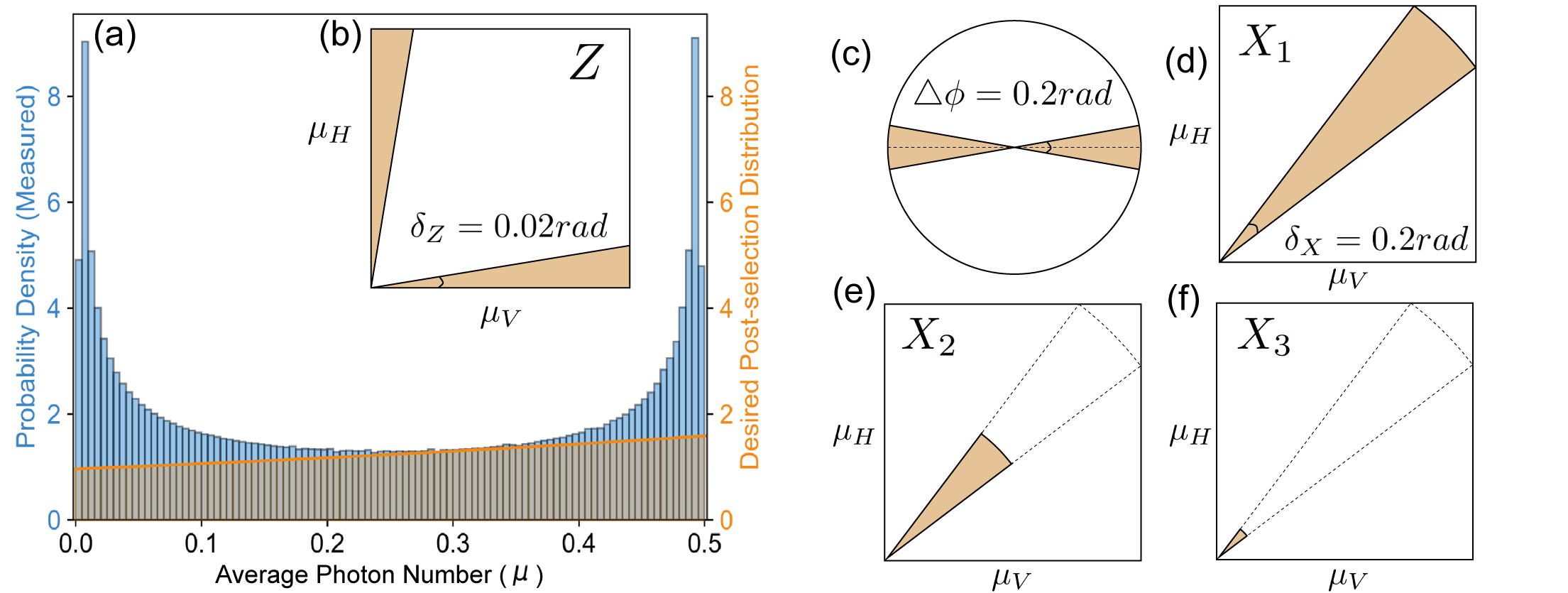}
		\caption{\textbf{Illustration of the post-selection protocol.} (a). The U-shape distribution of pulse intensities measured by PD1 (blue histogram) and the desired intensity distribution after post-selection (shaded area). (b). The region that defines Z-base data in the ($\mu_{H}$, $\mu_V$) plane, where the $\mu_{H}$ ($\mu_{V}$) denotes the intensity of H (V) component measured by PD1. (c). X-basis data (shaded area) defined on the equator of the Bloch sphere. (d), (e), (f) three different intensity sets ($X_{1}, X_{2}, X_{3}$) of X-base data are defined on ($\mu_{H}$, $\mu_V$) plane for decoy-state analysis.}
		\label{fig:3}
	\end{figure*}

	\textbf{Fully passive modulation.} As illustrated in Fig.~1, our fully passive modulation module takes four coherent states with identical intensities and random phases as inputs to produce any polarization state on the Bloch sphere with a random intensity between 0 and the maximum value. Generally, we can write the four strong inputs  as \{$\sqrt{\mu} e^{\phi_{1}}$,$\sqrt{\mu} e^{\phi_{2}}$,$\sqrt{\mu} e^{\phi_{3}}$,$\sqrt{\mu} e^{\phi_{4}}$\} and the whole process can be described as follows:

	(1) Passive decoy state generation. We divide the four input states of strong light into two groups and inject them into 50:50 beamsplitters for interference. Taking one of the groups as an example, we choose one of the interference outputs for local measurement and keep it for local measurement, while the other output serves as the H (V) polarization component that will be used for the next step. The intensities of the two polarization components can be expressed as $\mu_{H(V)}=\mu[1+\cos\phi_{H(V)}]$, in which $\phi_{H(V)}=\phi_{1(3)}-\phi_{2(4)}$, $\mu_{H(V)}\in [0,\mu_{max}]$. Notice that the phase $\phi_{i} (i=1,2,3,4)$ here are independent random values uniformly distributed in $[0,2\pi]$, the obtained intensities of H(V) components follow a U-shaped distribution, whose probability density function is expressed as $f(\mu)=1/\pi\sqrt{\mu(\mu_{max}-\mu)}$. Then a post-selection strategy will be used to reshape the intensity distribution to $g(\mu)=Ce^{u}$, where $C$ is a constant.

	(2) Passive polarization encoding. We obtain the final polarization-encoded state by combining the H and V components from the previous step through a polarization beam splitter. We also need to use a beamsplitter to split a portion of the strong light for local measurement. The obtained polarization state can be described by:

	\begin{equation}
	\left|\psi\right>=\cos(\frac{\theta}{2})\left|H\right>+\sin(\frac{\theta}{2})e^{i\phi}\left|V\right>
	\end{equation}
where $\cos(\theta/2)=\sqrt{\mu_{H}/(\mu_{H}+\mu_{V})}$ and $\phi=\phi_{V}-\phi_{H}$. It is not difficult to see from equation (1) that for any above-given states $\{\phi_{1},\phi_{2},\phi_{3},\phi_{4}\}$, the output state corresponds to a point on the Bloch sphere that can be uniquely determined by the polar angle $\theta$ and azimuthal angle $\phi$. Actually, it has been proved that any point on the Bloch sphere can also be generated from a set of input states $\{\phi_{i}\}$,\cite{wang2022fully} thus establishing a one-to-one mapping between the input set and output states. 

	\textbf{Experimental setup.} As illustrated in Fig.~2, an experimental setup with a single laser source is designed for a proof-of-principle demonstration of a fully passive QKD protocol. On the Alice side, in order to generate the initial coherent states with random phases, we utilize a laser diode modeled EP1550-0-NLW and operate it in gain-switch mode. In this mode, by controlling the driving current, the diode emits a pulsed laser with random phases seeded from spontaneous emission. By implementing closed-loop temperature control and precisely setting the driving current, we set the laser's central wavelength at 1547.38nm. Using the periodic driving signals generated by an arbitrary waveform generator (AWG, Keysight-M8195), we obtained laser pulses with a repetition rate of 20MHz, an average power of 0.1mW, and an effective pulse width of 2ns. Then, we use an erbium-doped fiber amplifier (EDFA) to boost the laser power to about 15mW, ensuring that high-precision local measurements can still be conducted even after the losses incurred by various devices.

The amplified pulses pass through a circulator and enter an asymmetric Michelson interferometer (MZI), composed of a balanced beam splitter and Faraday mirrors (FM). The delay between the two arms of the MZI is set to be precisely equal to one pulse period (50ns), allowing for interference between consecutive pulses.  One output from the interference proceeds to subsequent processing, while the other output is directed back through the circulator and detected by a photodetector (PD1) for local measurement, which corresponds to the projection measurement onto the Z basis on the Bob side.

Then, we use a combination of a polarization controller and a polarizer to purify the polarization of the signal light. Subsequently, the pulses enter a polarization-maintaining beam splitter and are equally divided into two paths. We add a 100ns fiber delay to one of the paths, which is precisely equal to two pulse periods. Then, we perform polarization synthesis for each pulse with its next-to-nearest neighboring pulse using a polarization combiner. The resulting polarized pulses are split into two paths by a beam splitter. Most of the power is used for local polarization measurements that correspond to the projection measurement onto the X basis on the Bob side, while a small portion is further attenuated by the variable optical attenuator (VOA). The VOA is precisely set to ensure that the maximum intensity of the WCP of H (V) polarization is about 0.5 photons per pulse before entering the quantum channel.

For the local measurement, we use commercial InGaAs-biased photodetectors with a bandwidth of 5GHz and a rising time of 70ps. And a high-speed data acquisition card (modeled ADQ32) with the highest sampling rate of 5GSa/s and 12-bit resolution is used here to record all the local measurement results. On the Bob side, a standard BB84 decoding module is constructed using a beamsplitter and two sets of polarization measurements that implement projection onto Z and X bases. Four single photon detectors ( Qube series from IDQ)  are used here, of which the average efficiency is 10\% and the average dark count rate is 500Hz in free running mode. We also use a time tagger device modeled ID900 to collect all the single photon events for post-processing. We use an optical attenuator instead of a real single-mode fiber spool to serve as the quantum channel for a proof-of-principle demonstration (see supplementary materials part I for more experimental details ).

	\textbf{Experimental results.} We characterize the gain-switch light source using the data detected by PD1 (in Fig.~2). And these data are also used as the local measurement for the Z basis, recording all the intensity information of H(V) components denoted by $\mu_{H}$($\mu_{V}$). As shown in Fig.~3(a), the histogram shows the statistical analysis of the output intensities of $2\times 10^{10}$ pulses (re-scaled to the range of [ 0, 0.5 ]), revealing a U-shaped distribution, which indicates that the pulse phases are random values uniformly distributed between 0 and $2\pi$\cite{yuan2014robust}. In order to decouple the distribution of polarization from intensities in decoy-state analysis, we need to randomly discard some of the pulses according to a specific strategy to obtain our desired distribution $g(\mu)=Ce^{u}$ (as shown in the shaded area of Fig.~3(a), see supplementary materials part II for details).

	\begin{table}[bth]
	    \centering
	    \caption{Parameters of Z-basis measurement under different channel loss.}
	    \begin{ruledtabular}
	    \begin{tabular}{cccc}
	        & 7.2dB & 11.6dB& 16.7dB \\
	        \colrule
	        $E_{\mu}(\%)$ & $2.13\pm0.0026$ & $2.27\pm0.0027$ & $2.30\pm0.0027$\\
	        $Q_{\mu}(10^{-3})$ & $9.24\pm0.017$ & $2.95\pm0.009$ & $0.95\pm0.005$ \\
	    \end{tabular}
	    \end{ruledtabular}
	\end{table}

	\begin{table}[thb]
	    \centering
	    \caption{Parameters of X-basis measurement under the channel loss of 16.7dB.}
	    \begin{ruledtabular}
	    \begin{tabular}{cccc}
	        & $X_{1}$ & $X_{2}$& $X_{3}$ \\
	        \colrule
	        $E_{\mu}(\%)$ & $1.49\pm0.001$ & $1.61\pm0.003$ & $2.12\pm0.190$\\
	        $Q_{\mu}(10^{-3})$ & $1.40\pm0.003$ & $0.61\pm0.006$ & $0.11\pm0.010$ \\
	    \end{tabular}
	    \end{ruledtabular}
	\end{table}
In this work, we only use the Z-basis data for key generation. Unlike the active QKD, we use a region near the poles of the Bloch sphere that can be defined by $\{\mu_{H},\mu_{V}\}$ to determine the Z basis, denoted by the colored region in Fig.~3(b). We set the tolerance to be $\delta_{Z}=0.02rad$, where $\delta_{Z}$ can be calculated using $\tan(\delta_{Z})=\mu_{H}/\mu_{V}$. The determination of the X basis is slightly more complex. First, we need to use$\{\mu_{H},\mu_{V}\}$ to filter out pulses located near the equator of the Bloch sphere, which can be denoted by the colored region in Fig.~3(d). The tolerance is set to be $\delta_{X}=0.2$, which can be calculate using $\tan(\pi/4\pm\delta_{X}/2)=\mu_{H}/\mu_{V}$. Based on that, we further filter out pulses within a specific azimuthal angle range $\triangle\phi=0.2$ as shown in Fig.~3(c), to obtain the diagonal (antidiagonal) polarization states D(A). To do that, we need to combine all the local measurement results. We denote the measurement results of PD2 as $\mu_{D}$ and calculate the azimuthal angle $\phi$ by:
	\begin{equation}
	\phi=\arccos(\frac{\mu_{D}-\mu_{A}}{2\sqrt{\mu_{H}\mu_{V}}})
	\end{equation}
where $\mu_{A}$ can be calculated with $\{\mu_{H}, \mu_{V}, \mu_{D}\}$ .
As shown in Fig.~3(c), we define the region $\phi=\pm0.1rad$ ($\phi=\pi\pm0.1 rad$) as D (A) states.

It is worth noting that in Fig.~3(d), the selected region is a sector rather than a quadrilateral that includes the top-right corner, which is for the convenience of decoy-state analysis (see supplementary materials part III). We represent this region as X1. Similarly, we represent the other two sector-shaped subregions of X1 as X2 and X3 (shown in Fig.~3(e)(f)), with their radii being 0.5 and 0.1 of X1, respectively. By utilizing these three different intensity sets, we then are able to do the decoy-state analysis\cite{ma2005practical, hwang2003quantum, wang2005beating}.

We conduct QKD experiments at three different channel losses: 7.2dB, 11.6dB, and 16.7dB. For each loss situation, we collect data for 1000s with a total data size of $2\times 10^{10}$. Table.~1 lists the error rate $E_{u}$ and gain $Q_{u}$ measured in three different channel losses for Z basis. For the X basis, we analyze data from three different intensity regions \{$X_{1}, X_{2}, X_{3}$\} under each loss scenario. Here, the parameters for a 16.7 dB loss scenario are presented in Table 2 and other parameters can be found in supplementary materials part IV. Based on these measured results, we calculate the final key rate using the following formula\cite{wang2022fully}:
	\begin{equation}
	\begin{aligned}
	R = P_{Z}\{ \langle P_{1} \rangle_{S_{Z}}Y_{1}^{Z,perfect,L}[1-h_{2}(e_{1}^{X,perfect,U})]\\
-f_{e}\langle Q_{Z} \rangle_{S_{Z}} h_{2}(\langle QE_{Z} \rangle_{S_{Z}}/\langle Q_{Z} \rangle_{S_{Z}})\}
	\end{aligned}
	\end{equation}
where $h_{2}(x)=-x\log_{2}(x)-(1-x)\log_{2}(1-x)$ is the binary entropy function. As plotted in Fig.~4, the obtained final key rate are $7.62\times10^{-5}$ in 7.2dB, $4.01\times10^{-5}$ in 11.6dB and $1.78\times10^{-6}$ in 16.7dB in the asymptotic regime.

	\begin{figure}[htb]
		\includegraphics[scale=0.6]{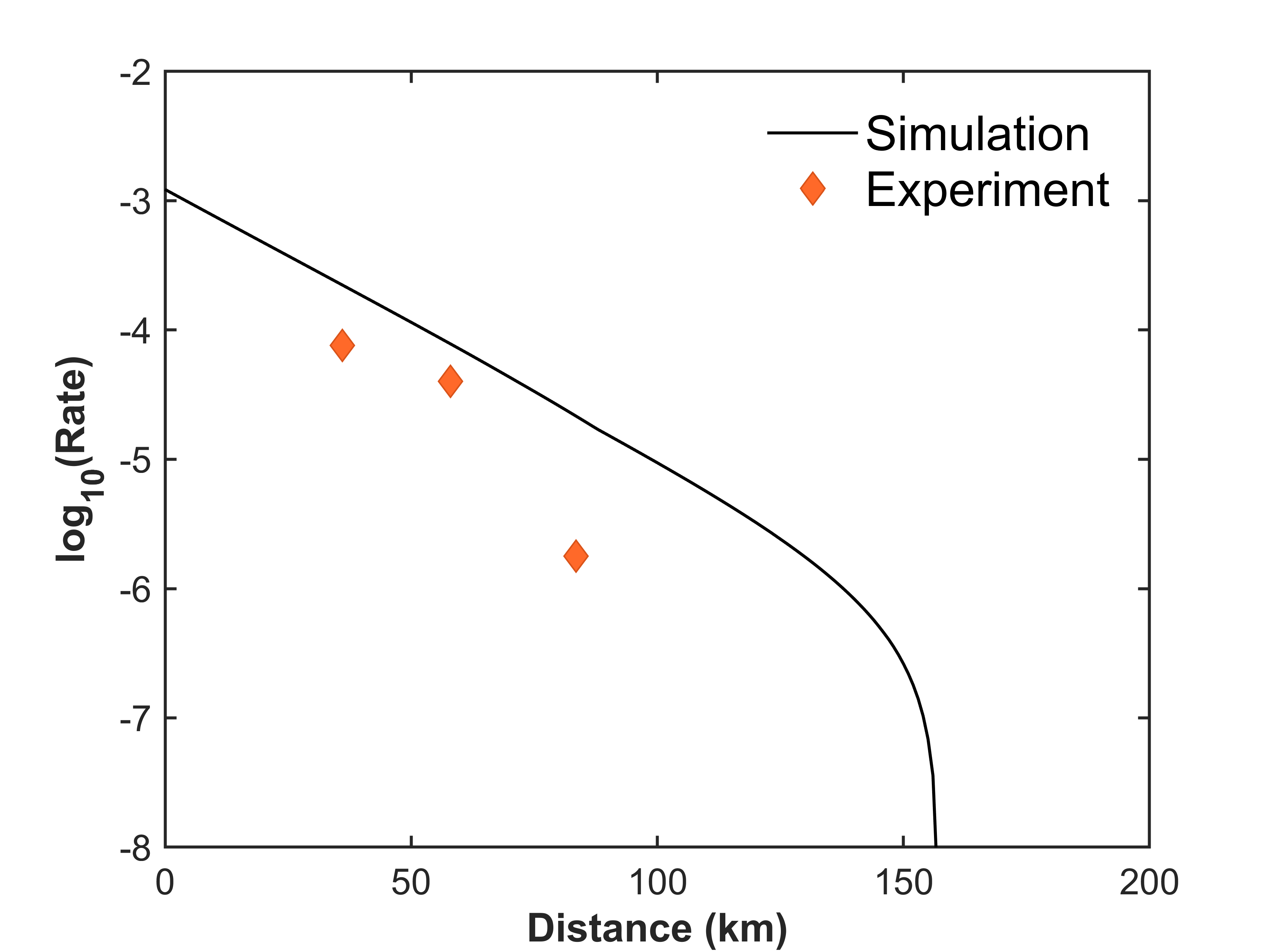}
		\caption{\textbf{Final key rate simulation and experimental results.} The line in the graph represents the simulation of the key rate at different distances, assuming that the channel is a standard telecom fiber with an attenuation of 0.2 dB/km. Other parameters include the single-photon detection efficiency of 10\% and the dark count rate of $10^{-6}$.}
		\label{fig:4}
	\end{figure}

	\begin{figure}[htb]
		\includegraphics[scale=0.98]{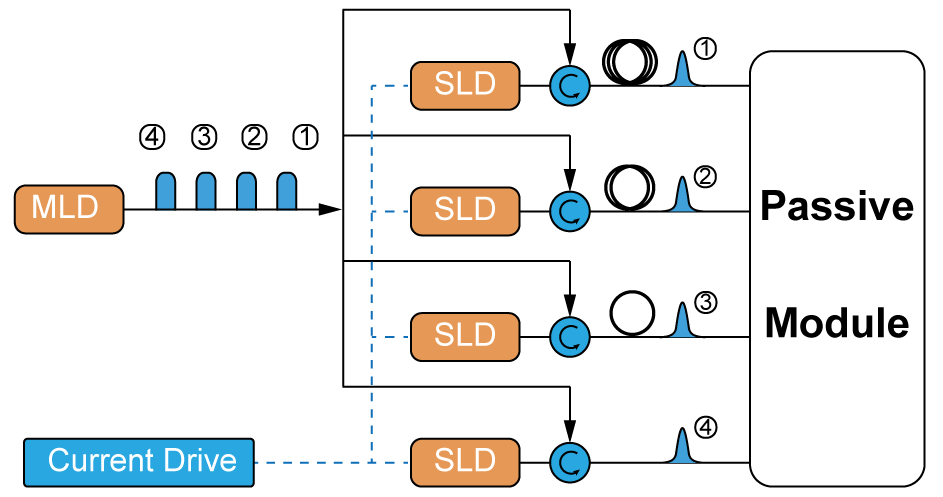}
		\caption{\textbf{Fully passive source with injection locking.} When the first seed pulse arrives, the drive current of the first slave laser is activated. Similarly, when the second seed pulse arrives, the drive current of the second laser is activated, and so on. MLD: master laser, SLD: slave laser.}
		\label{fig:4}
	\end{figure}

	\textbf{Discussion} In summary, we successfully demonstrated fully passive quantum key distribution using a single laser source with polarization encoding, under different channel losses, which completely eliminates the side channels introduced by active modulation.

Compared to multi-laser schemes, using a single laser can avoid wavelength side channels and prevent the decrease in interference visibility caused by frequency deviations\cite{moschandreou2018experimental, yin2016measurement}. However, since we use interference between early-late pulse pairs, the output WCPs sequence will not be completely independent of each other. In our protocol, only the first pulse out of every four can be used as a quantum signal, while the remaining three need to be discarded (as a proof-of-principle demonstration, we discard these pulses through post-selection). One solution to this would be adding extra intensity modulators or optical switches to pick the valid pulse. Notice that those modulators or optical switches are only used to periodically switch the signal on and off, without involving any modulation information, so it does not generate side channels. 

Here, we present a more elegant proposal: one can combine injection locking techniques and fully passive modulation to achieve a completely modulator-free scheme\cite{lo2023simplified}. As illustrated in Fig.~5, the master laser generates phase-randomized pulses in gain-switch mode, grouping these pulses in sets of four. By applying driving currents at appropriate timings, the four slave lasers are each seeded by one of the four pulses, thus inheriting their respective phase information. After a suitable delay, the pulses generated by the slave lasers are injected into a fully passive modulation system, thereby resolving the issue of non-independence between adjacent pulses.

\begin{acknowledgments} \textbf{Acknowledgments.} We thank Li Qian, Zhiliang Yuan, Gai Zhou, Xiong Wu, Rong Wang, and Chenyang Li for their helpful discussions. The project is supported by the University of Hong Kong start-up grant. W. Wang acknowledges support from the Hong Kong RGC General Research Fund (GRF) and the University of Hong Kong Seed Fund for Basic Research for New Staff. H.K. Lo is also supported by NSERC, MITACS, and Innovative Solutions Canada.

\end{acknowledgments}

\textit{Note added.} Recently, we become aware of a related work\cite{lu2023experimental}.
	
	
%

\end{document}


\title{Proof-of-Principle Demonstration of Fully-Passive Quantum Key Distribution: Supplementary Materials}

	\author{Chengqiu Hu}
	\thanks{hucq@hku.hk}
	\affiliation{Department of Physics, University of Hong Kong, Pokfulam Road, Hong Kong}
	
	\author{Wenyuan Wang}
	\affiliation{Department of Physics, University of Hong Kong, Pokfulam Road, Hong Kong}
	
	\author{Kai-Sum Chan}
	\affiliation{Department of Physics, University of Hong Kong, Pokfulam Road, Hong Kong}
	\affiliation{Quantum Bridge Technologies, Inc., 100 College Street, Toronto, ON M5G 1L5, Canada}
	
	\author{Zhenghan Yuan}
	\affiliation{Department of Physics, University of Hong Kong, Pokfulam Road, Hong Kong}

	\author{Hoi-Kwong Lo}
	\thanks{hoikwong@hku.hk}
	\affiliation{Department of Physics, University of Hong Kong, Pokfulam Road, Hong Kong}
	\affiliation{Centre for Quantum Information and Quantum Control (CQIQC), Dept. of Electrical \& Computer Engineering, University of Toronto, Toronto,  Ontario, M5S 3G4, Canada}
	\affiliation{Centre for Quantum Information and Quantum Control (CQIQC), Dept. of Physics, University of Toronto, Toronto,  Ontario, M5S 3G4, Canada}
	\affiliation{Quantum Bridge Technologies, Inc., 100 College Street, Toronto, ON M5G 1L5, Canada}

	\date{\today}
	\maketitle

	\section{\romannumeral1. Experimental Details}
In this part, we will present more experimental details of passive modulation. As shown in Fig.~1, the passive modulation module consists of two parts connected by a curved arrow. In the first part, we use a gain-switch laser to generate a sequence of pulses at 20MHz (indicated on the right of the figure using 1, 2, ..., 10, ...). We utilize an asymmetric Michelson interferometer to interfere the original pulse sequence with a time delay of one pulse T, resulting in a new pulse sequence (represented as 1+2, 2+3, 3+4, ...). Due to the random phase of the original pulses, the intensity of the new pulse sequence is also random. We can obtain the intensity information of all the pulses (energy conservation) by measuring with PD1. In the second part, we divide the polarization-purified pulse sequence into two paths with a PMBS, labeled as H and V, respectively. The H path is delayed by two pulse positions (2T) and then recombined with the V path using a PBC. In this way, we obtain the final output pulse sequence, represented as (1234), (2345), etc. From the labeling of these pulses, it can be observed that not all pulses are independent of each other. In order to eliminate the correlation between pulses, we discarded some pulses through post-selection, such as (2345), (3456), (4567), etc. The pulses that were preserved are indicated by red boxes in the figure.
\begin{figure*}[htb!]
		\includegraphics[scale=0.98]{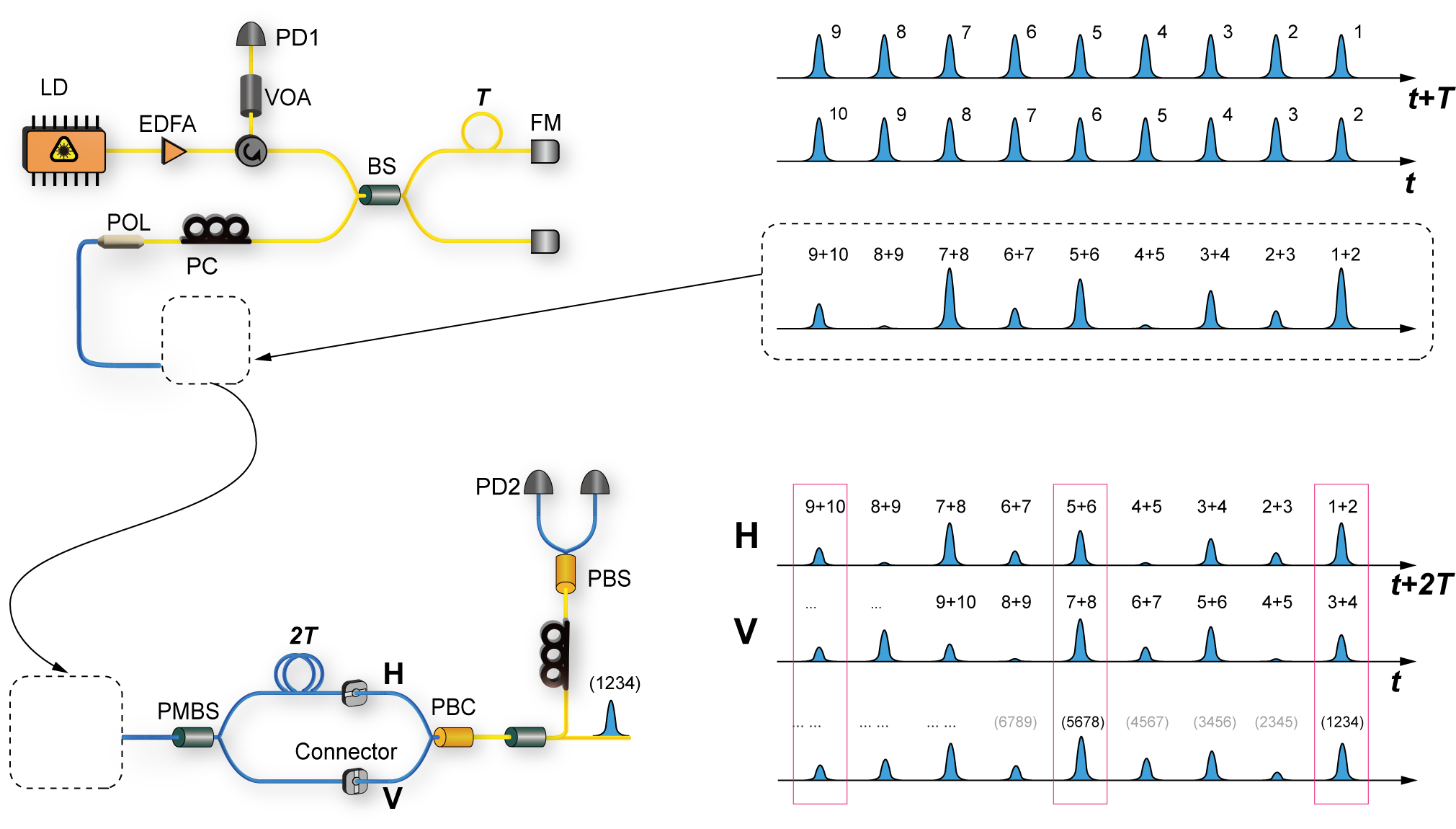}
		\caption{\textbf{Illustration of fully passive modulation.}}
		\label{fig:S1}
	\end{figure*}

Next, we will explain how to perform local measurements using the data from PD1 and PD2. As shown in Fig.~1, the H and V components are obtained from the same pulse sequence through time-multiplexing. All the intensity information of this pulse sequence can be obtained from the detection of PD1. Let us denote the intensity of pulse 1+2 (3+4) in Fig.~1 with $\mu_{H}$ ($\mu_{V}$). Therefore, the detection of PD1 is equivalent to a Z-basis local measurement. Similarly, the detection of PD2 corresponds to an X-basis local measurement, which requires adjusting the polarization controller to align the projection direction here with the X-basis projection direction at the Bob side. Let us denote the intensity detected by PD2 with $\mu_{D}$. Notice that we use a trick here to minimize the data collection for local measurements. By adjusting the variable optical attenuator (VOA) in front of PD1, we can make sure that $max(\mu_{H})+max(\mu_{V})=max(\mu_{D})$. As shown in Fig.~2, the blue dots represent the detection data of PD2, and the orange dots represent the detection data of PD1. In this way, we do not need an additional detector to obtain the data for $\mu_{A}$. Through energy conservation, we have $\mu_{H}+\mu_{V}=\mu_{D}+\mu_{A}$.

\begin{figure}[htb!]
		\includegraphics[scale=0.6]{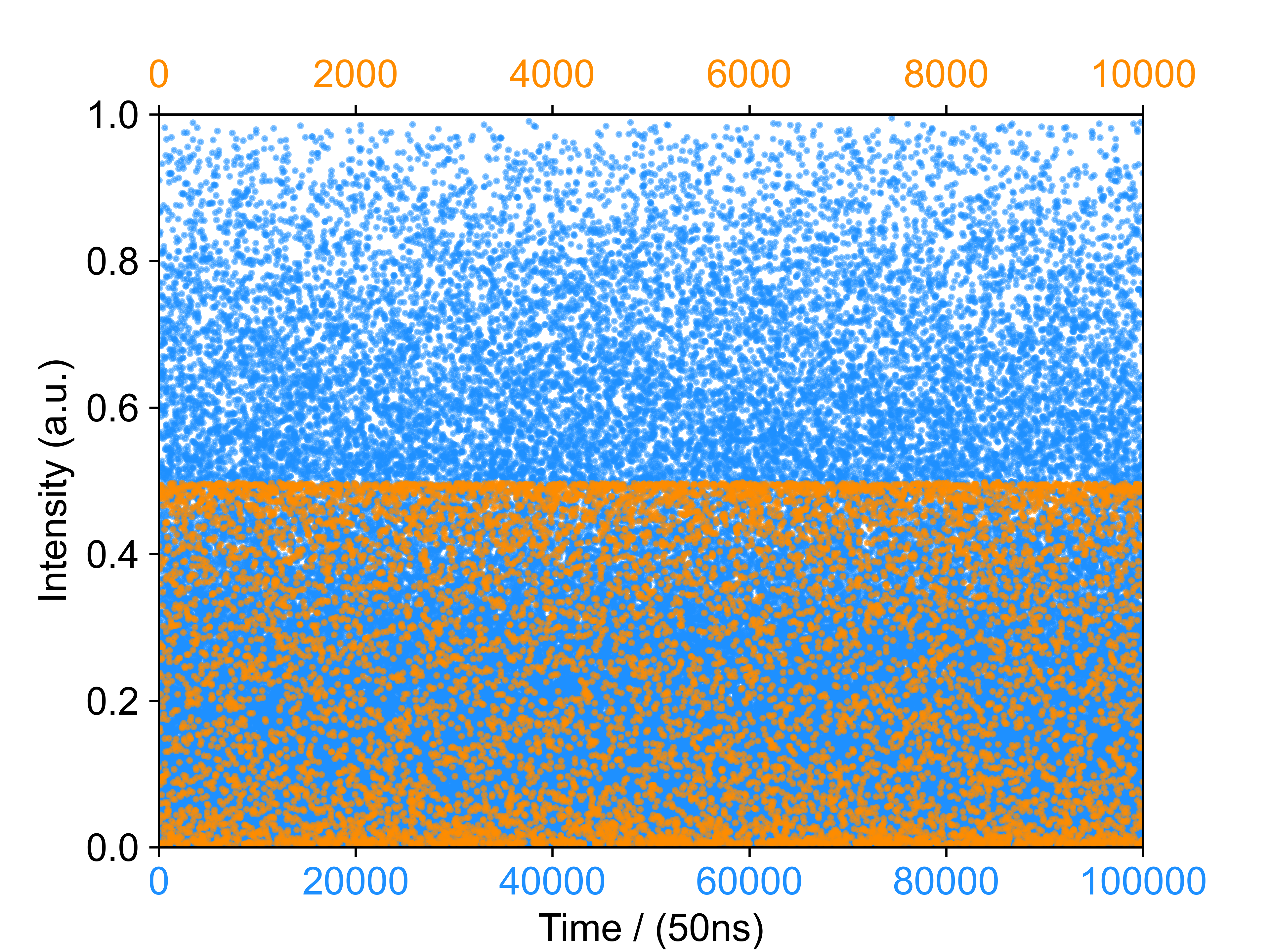}
		\caption{\textbf{Results of local measurement.}}
		\label{fig:S2}
	\end{figure}

	\section{\romannumeral2. Post-selection strategy}
In this part, we will introduce the detailed post-selection strategy in our experiment. In order to get the desired exponential distribution (denoted by $g(\mu_{H/V})$) from the original distribution of the signal intensities, we use the so-call Acceptance-Rejection method. By precisely adjusting the VOA at the Alice end, we attenuate the maximum values of the output quantum signals to $max(\mu_{D/A})=1$ and $max(\mu_{H/V})=0.5$, respectively. Therefore, we need to scale all the local measurement data to the range of $[ 0, 0.5 ]$ before further processing (shown in Fig.~2). Then we obtain the distribution of the original intensities by separating all the data into 100 bins and calculating their probability density values, which can be approximately fitted with a U-shaped curve described by $f(\mu)=1/\pi\sqrt{\mu(0.5-\mu)}$. We define $g(\mu_{H/V})=C\exp(\mu_{H/V})$,  where the constant $C$ is the maximum value that satisfies the condition $C\exp(\mu_{H/V})<=f(\mu_{H/V})$ for all $\mu_{H/V}$ ($C=0.961$ in our case). For each original data $\mu_{H/V}$, we generate a uniform random number $x \in [ 0, 1 ]$ and keep $\mu_{H/V}$ as effective data if $x<g(\mu_{H/V})/f(\mu_{H/V})$. After that, about $38.8\%$ of the original signals are kept and we then obtain the exponential distribution of signal intensities.

Alice can then define regions on $(\mu_H,\mu_V,\phi)$ (or, more conveniently, two-dimensional regions on $(\mu_H,\mu_V)$ and one-dimensional regions on $\phi$, which are decoupled from each other) to further post-select signals and define her states conditionally. The signals are first divided by Alice into testing and key-generation datasets. For the infinite-data scenario (which this work focuses on) we can assume that only an infinitely small portion of the data is used for testing, and the proportion of key-generation dataset approaches one, i.e. the division between testing and key-generating datasets does not affect the sifting efficiency. For the key-generating dataset, Alice only needs to prepare the Z basis. She can define two triangular regions (as shown in the main text Fig. 3 (b)) to represent her H and V states. The regions are defined by the parameter $\delta_Z$. The 2-D polar angles on the $(\mu_H,\mu_V)$ map for Alice's H and V states are respectively $[0,\delta_Z]$ and $[\pi/2 - \delta_Z, \pi/2]$ ($\delta_Z=0.02rad$ in our case).

For the testing (decoy-state) dataset, Alice needs to prepare states in both X and Z bases, as well as in three decoy intensities. This is achieved by her selecting three sector-shaped regions with polar angles $[0,\delta_Z]$, $[\pi/2 - \delta_Z, \pi/2]$ and $[\pi/4 - \delta_{X}, \pi/4 + \delta_{X}]$, respectively representing the H, V and the diagonal states $\{+,-\}$ (which all fall on the equator of the Bloch sphere). Each sector shape intersects with the boundaries $\mu_{max}$ and has a maximum radius $r_{max}$. The sectors are then scaled by factors $t_i\in \{t_{decoy2},t_{tecoy},1\}$ to generate the three decoy settings with intensities ranging in $[0,t_i \times r_{max}]$, which take the shapes of concentric sectors \footnote{There is overlap between them, but this is valid since the linear constraints in decoy analysis can be freely linearly combined so long as the total number of independent equations stays the same.}. Example X basis regions are shown in the main text Fig. 3 (d) - (f).

More details on the post-selection regions, especially a full numerical description, can be found in Ref. \cite{passiveQKD} Supplemental Materials.

	\section{\romannumeral3. Decoy-State Analysis and Key Rate Formula}

In this part. We will briefly recapitulate the decoy-state analysis, as well as the key rate formula. In Ref. \cite{passiveQKD}, it is proven that one can write the decoy constraints as:

\begin{equation}
\begin{aligned}
	\langle Q \rangle_{S_i} &= \sum_{n} \langle P_n \rangle_{S_i} Y_n^{mixed}\\
	\langle QE \rangle_{S_i} &= \sum_{n} \langle P_n \rangle_{S_i} e_n Y_n^{mixed}\\
\end{aligned}
\end{equation}

\noindent where $Y_n^{mixed}$ and $e_nY_n^{mixed}$ are the yield and error-yield of $n$-photon states with mixed polarizations (resulting from the polarization distribution in a given post-selection region $S_i$) \footnote{Note that a region can include phase-randomized coherent states with various intensities and polarizations. While the intensities do not affect the Fock state statistics, the polarizations will. Therefore, $Y_n^{mixed}$ and $e_nY_n^{mixed}$ are the yield and error-yield averaged over the polarization distribution in a given region $S_i$.} and $\langle f(\mu_H,\mu_V,\phi_{HV}) \rangle$ is the expected value of a given observable over $S_i$, i.e. 

\begin{equation}
	\langle f(\mu_H,\mu_V,\phi_{HV}) \rangle = {{\iiint_{S_i} f(\mu_H,\mu_V,\phi_{HV}) p(\mu_H,\mu_V,\phi_{HV}) d\mu_H d\mu_V d\phi_{HV}}\over{\iiint_{S_i} p(\mu_H,\mu_V,\phi_{HV}) \rangle d\mu_H d\mu_V d\phi_{HV}}}
\end{equation}
In the experiment, the observables $\langle Q \rangle_{S_i}$ and $\langle QE \rangle_{S_i}$ correspond to actual number of signals that satisfy the criteria for a given post-selection region $S_i$, divided by the total number of signals.

One can use the above decoy constraints in a linear program in the same way as what one does for active QKD, from which we can obtain the upper and lower bounded values of single-photon statistics, $Y_1^{mixed,L}$ and $e_1Y_1^{mixed,U}$ (and by extension $e_1^{mixed,U}=e_1Y_1^{mixed,U}/Y_1^{mixed,L}$).

Note that, since in the decoy analysis we calculate the yields of a given basis (and not a given state), $S_i$ above actually can correspond to pairs of regions (e.g. both H polarization states and V polarization states with the same intensity range $[0,t_i \times r_{max}]$). For simplicity we omit the $X$ and $Z$ basis superscripts, but note that we can use the same form of the equations to obtain e.g. both $Y_1^{X,mixed,L}$ and $Y_1^{Z,mixed,L}$.

It is further proven in Ref. \cite{passiveQKD} that these upper and lower bounds are also valid upper and lower bounds for perfectly prepared (i.e. pure state) single photons in e.g. $H,V,+,-$ polarizations. In other words,

\begin{equation}
	\begin{aligned}
		Y_1^{Z,perfect} &\geq Y_1^{Z,mixed,L} \\
		e_1^{X,perfect} &\leq e_1^{X,mixed,U} \\
	\end{aligned}
\end{equation}
\noindent so we can use these bounds on mixed states directly in the same key rate formula for active BB84 (which should have been based on the scenario where Alice prepares pure states). The key rate can be bounded by

\begin{equation}
	R \geq \langle P_Z \rangle_{S_{Z,key}} \left\{\langle P_1 \rangle_{S_{Z,key}} Y_1^{Z,mixed,L} [1-h_2(e_1^{X,mixed,L})] - f_e \langle Q \rangle_{S_{Z,key}} h_2 (\langle QE \rangle_{S_{Z,key}} / \langle Q \rangle_{S_{Z,key}})\right\}
\end{equation}

\noindent where $S_{Z,key}$ is the key-generating Z basis region (i.e. summing over both H and V states), $P_1$ is the Poissonian distribution for single photons $P_1=(\mu_H+\mu_V) e^{\mu_H+\mu_V}$ and $h_2$ is the binary entropy function.

Again, a more detailed discussion on the decoy-state analysis (and why it is valid) can be found in Ref. \cite{passiveQKD}.

	\section{\romannumeral4. Detailed Experimental Results}
In this part, we show more detailed experimental results. Table.~I and Table.~II list the parameters for X basis under channel loss of 7.2dB and 11.6dB.
	\begin{table}[thb]
	    \centering
	    \caption{Parameters of X-base measurement under a channel loss of 7.2dB..}
	    \begin{ruledtabular}
	    \begin{tabular}{cccc}
	        & $X_{1}$ & $X_{2}$& $X_{3}$ \\
	        \colrule
	        $E_{\mu}(\%)$ & $1.36\pm0.001$ & $1.42\pm0.003$ & $1.68\pm0.016$\\
	        $Q_{\mu}(10^{-2})$ & $1.39\pm0.001$ & $0.63\pm0.002$ & $0.12\pm0.004$ \\
	    \end{tabular}
	    \end{ruledtabular}
	\end{table}

	\begin{table}[thb]
	    \centering
	    \caption{Parameters of X-base measurement under a channel loss of 11.6dB..}
	    \begin{ruledtabular}
	    \begin{tabular}{cccc}
	        & $X_{1}$ & $X_{2}$& $X_{3}$ \\
	        \colrule
	        $E_{\mu}(\%)$ & $1.36\pm0.001$ & $1.45\pm0.003$ & $2.04\pm0.018$\\
	        $Q_{\mu}(10^{-3})$ & $4.39\pm0.006$ & $2.02\pm0.012$ & $0.33\pm0.024$ \\
	    \end{tabular}
	    \end{ruledtabular}
	\end{table}

	
		

